# On the detection of finite-frequency current fluctuations

G. B. Lesovik

*Institute of Solid State Physics, Russian Academy of Sciences, 142432 Chernogolovka, Moscow Region, Russia*

R. Loosen

*Bayer AG, 51368 Leverkusen, Germany*

10 February 1997

We consider a measurement of finite-frequency current fluctuations, using a resonance circuit as a model for the detector. We arrive at an expression for the measurable response in terms of the current–current correlators which differs from the standard (symmetrized) formula. The possibility of detection of vacuum fluctuations is discussed.

Finite-frequency (FF) current fluctuations at zero temperature (vacuum fluctuations, VFs) have been discussed for a long time[1] in connection with the analogous question of electromagnetic vacuum fluctuations. Recently there has been renewed interest in the noise at finite frequency in connection with the supposed possibility of observing the Fermi edge singularity in noninteracting[1,2] and interacting[3] systems.

In the present letter we consider a realistic model for an FF measurement and show that, in a very close analogy with the electromagnetic vacuum fluctuations, a certain measurability limitation appears.

There are different practical and theoretical approaches to FF measurements:

1. Making repeated measurements of the instantaneous values of the current over a long time interval and later Fourier transforming the data obtained.

2. Making a single measurement of the charge transmitted during a given time interval. In that case the information about the FF fluctuation appears through an integral over all frequencies. Ideally that can be done by making two measurements of the charge in the reservoir, the initial (during system preparation) and final. An alternative measurement can be made with a ''Larmor clock'' (the spin rotating in the magnetic field produced by the current); this method, which is described in Ref. 4, can perhaps be implemented.

3. Making a time-averaged measurement of the response of a resonance circuit, which can be an ordinary $LC$ element, i.e., an inductive element coupled to the quantum wire, a capacitor whose charge is the quantity to be measured as a response, and the resistance of the circuit.

The last approach, we believe, is the most relevant for FF measurements.

We model our detector (the resonator, which we will refer to as $LC$) by an oscillator and consider the response in the first nonvanishing order in the inductive coupling constant (we will outline the calculation at the end).

Finally, we arrive at the practical conclusion that the measurable response of such a model detector at a certain frequency $\Omega$ can be written in terms of the usual current–current correlators as follows:

$$S_{\text{meas}} = K\{S_+(\Omega) + N_\Omega[S_+(\Omega) - S_-(\Omega)]\}, \tag{1}$$

with the definitions

$$S_+(\Omega) = \int dt \langle I(0)I(t) \rangle \exp(i\Omega t),$$

and

$$S_-(\Omega) = \int dt \langle I(t)I(0) \rangle \exp(i\Omega t).$$

The frequency $\Omega$ in all the expressions is strictly positive. $N_\Omega$ stands for the Bose occupation number of the oscillator, i.e., $N_\Omega = [\exp(\hbar\Omega/k_B T_{LC}) - 1]^{-1}$, $K$ is the effective coupling constant of the quantum wire with the resonator, $\langle A \rangle = \text{Tr}\rho A$, where $\rho$ is the density matrix of the electrons, and the time-dependent current operators are defined in the standard way: $I(t) = \exp(iHt) I \exp(-iHt)$.

The expression obtained here should be contrasted with the widely used formula[5]

$$S(\omega) = \int dt \, \exp(i\omega t) \left\langle \frac{1}{2}\{I(0)I(t) + I(t)I(0)\} \right\rangle. \tag{2}$$

Note that this formula contains the symmetrized current–current correlator. The necessity of such a symmetrization comes from the fact that in the general case the current operators at different times do not commute with each other, and it is motivated by the close correspondence with the classical expression.[5]

Using the definition (2), one arrives at the well-known expression for the spectral density of the current–current correlator for a conductor in equilibrium:[6]

$$S(\omega) = 2G\hbar\Omega \left[ \frac{1}{2} + \frac{1}{\exp(\hbar\Omega/k_B T) - 1} \right]. \tag{3}$$

This expression tells us that at zero temperature one should expect fluctuations proportional to the frequency, which are interpreted as an analog of the vacuum fluctuations of the electromagnetic field.

Nevertheless, as is known from optical measurements, ordinary photodetectors do not react to the vacuum fluctuations, because it is not possible to take energy from the vacuum to excite atoms in the detector.[7]

Yet the vacuum fluctuations are observable, though less directly than are ordinary fluctuations, via the Lamb shift[8] or the Casimir effect[9] or by using a so-called quantum photocounter,[10] which is prepared in an excited state and can thus react to VFs.

As we will now show by analyzing Eq. (1), an $LC$ detector may operate as a photodetector without any reaction to VFs or as an optical detector for VFs, but it never gives the standard Nyquist expression for FF noise (3), as one might naively expect.

If the detected frequency is higher than the $LC$ temperature, the occupation $N_\Omega$ is exponentially small, and the only nonvanishing term in (1) is the ''positive part'' of the spectral density $S_+(\Omega)$, which describes the ''emission'' of energy by the conductor to the $LC$ circuit, and in that case the $LC$ circuit functions as an ordinary photodetector.

As an example, for $S_+(\Omega)$ in a coherent conductor with transmission $D$ at zero temperature and a finite bias voltage one has

$$S_+(\Omega) = \frac{2e^2}{h} D(1-D)(eV - \hbar\Omega) \tag{4}$$

if $\hbar\Omega < eV$ and $S_+(\Omega) = 0$ otherwise.

We have neglected the energy dependence of the transmission in the expression above, as well as an additional frequency dependence which has to come from the averaging over the coordinates (see below). Equation (4) coincides with the *excess* spectral density calculated in Ref. 1 using the symmetrized correlator (2).

If the frequency is much lower than the detector temperature, $\hbar\Omega \ll k_B T_{LC}$, one may replace the Bose occupation number $N_\Omega$ by $k_B T_{LC}/\hbar\Omega$.

The difference $S_+(\Omega) - S_-(\Omega)$ is *negative*, and in the case of a quantum conductor, provided that the transmissions $D_n$ depend only weakly on the energy we find

$$S_+ - S_- = -2\hbar\Omega G, \tag{5}$$

where $G = 2e^2/h \sum_n D_n$ is the conductance.[11]

Note that the singular behavior of the spectral density at $\hbar\Omega = eV$ which was found in Ref. 2 for the symmetrized expression $S_+ + S_-$ is not present in $S_+ - S_-$, and we conclude that the measurable singularity at zero temperature and finite bias is due solely to the cutoff of the frequency by the voltage in $S_+(\Omega)$ (4).

Altogether, for $\hbar\Omega \ll k_B T_{LC}$ we have

$$S_{\text{meas}} = K\{S_+(\Omega) - 2Gk_B T_{LC}\}. \tag{6}$$

The meaning of the negative part is clear—the $LC$ circuit is ''cooled down'' by emitting energy into the conductor. So, in some sense, in this limit the vacuum fluctuations, represented by $S_-$, are detectable; note, however, that they appear in the answer in a way which is quite different from the Nyquist expression (3).

If the conductor is in equilibrium (aside from the weak interaction with the $LC$ circuit), for low frequencies we find

$$S_{\text{meas}} \propto 2G(T_e - T_{LC}). \tag{7}$$

This expression vanishes if the electron temperature $T_e$ is equal to the $LC$ temperature $T_{LC}$, as is expected for overall equilibrium.

At intermediate frequencies, where $k_B T_e, eV_{\text{bias}} \ll \hbar\Omega \ll k_B T_{LC}$ the measurable response is negative:

$$S_{\text{meas}} = -2Gk_BT_{LC}. \tag{8}$$

Let us now outline briefly the derivation of Eq. (1). Our chosen detector can be regarded as a harmonic oscillator coupled linearly to the time derivative of the current to be measured:

$$M\ddot{x}(t) = -Dx(t) - \alpha \dot{I}(t). \tag{9}$$

In terms of the physical parameters of the counter (inductance $L$, inductive coupling, and capacitance $C$), we may write: $M = L$, $D = 1/C$, resonance frequency $\Omega = \sqrt{D/M} = \sqrt{1/LC}$, while $\alpha$ is the inductive coupling itself.

Our goal is to calculate the change in $x^2$ due to current fluctuations in the first nonvanishing order of perturbation theory.

We must evaluate the expression:

$$\langle x^2(0) \rangle = (-i\alpha)^2 \int_{-\infty}^{0} dt_1 \int_{-\infty}^{t_1} dt_2 e^{\eta(t_1+t_2)} \langle [[x^2(0), x(t_1)\dot{I}(t_1)], x(t_2)\dot{I}(t_2)]] \rangle. \tag{10}$$

The angle brackets here stand for averaging over the unperturbed density matrix of the electrons and the oscillator. Evaluating the expression above, we get

$$\langle x^2(0) \rangle = \left(\frac{\alpha}{2M\Omega}\right)^2 \int_{-\infty}^{+\infty} dt \, \exp(-\eta|t| + i\Omega t) \left[\frac{1}{\eta} - \frac{1}{\eta - i\,\text{sign}(t)\Omega}\right]$$

$$\times \{\langle \dot{I}(0)\dot{I}(t) \rangle (1 + N_\Omega) - \langle \dot{I}(t)\dot{I}(0) \rangle N_\Omega\}. \tag{11}$$

The derivation of Eq. (1) from this equation is not a straightforward procedure, the main problem being that the integral over frequencies $\omega$ of the Lorenzian $\eta/\eta^2+(\omega-\Omega)^2$, which is supposed to serve as a delta function, contains a factor of $\omega^2$ and does not converge well. If we neglect that problem and replace the Lorenzian by a delta function, $\eta/\eta^2+(\omega-\Omega)^2 = \pi\delta(\omega-\Omega)$, we end up with Eq. (1) with the coefficient $K$ in it given by

$$K = \left(\frac{\alpha}{2L}\right)^2 \frac{1}{2\eta}. \tag{12}$$

The coefficient diverges as the width of the resonance $\eta$ goes to zero, so we keep the latter small but finite.

The shape $F(\omega - \Omega)$ of the resonance in the $LC$ circuit, when calculated more exactly, is in fact not a Lorenzian, and without specifying it we may write instead of Eq. (1)

$$S_{\text{meas}} = \int \frac{d\omega}{2\pi} F(\omega - \Omega)\{S_+(\omega) + N_\Omega[S_+(\omega) - S_-(\omega)]\}. \tag{13}$$

The function $F(\Delta\omega)$ can in principle be measured independently, by applying an alternating current. Afterwards the measured expression can be substituted into the formula above.

In general the current operator and its average over the density-matrix correlators at different times depends on the coordinates. The operators $I(t)$ used in this paper are in fact the total current operators averaged, in addition, over the length $l$ of the inductive-coupling region,

$$I(t) \equiv 1/l \int_{X-l/2}^{X+l/2} I(r,t) dr.$$

Because of this symmetry with respect to the coordinates, the current operators stand as $I(r_1,t) + I(r_2,t)$, although we keep a certain order in time.

Apart from the symmetrization with respect to the coordinates, the problem of the ordering of the current operators stems from the presence of the vacuum fluctuation part. If discussion is limited to the low-frequency limit $\hbar\Omega \ll eV_{\text{bias}}, k_B T$, then it does not matter whether one uses $S_+(\Omega)$, $S_-(\Omega)$, or the Fourier-transformed symmetrized equation (2), the result will be the same up to small corrections of the order of $\hbar\Omega/eV$, $k_B T$.

We are indebted to Janosh Hajdu for many helpful discussions and for his participation in the initial stages of this work. We are also thankful to Christian Glattli for interesting discussions of the results and useful comments. The work of G. B. L. was done partly during his stay in Cologne ITP, Germany; Saclay, France; and ITP ETH Zürich, Switzerland and was supported by SFB 341 in Germany, the Swiss National Foundation, and in part by the Russian Academy of Sciences, RFFR Grant No. 96-02-19568.

[1)] J. Hajdu, private communication.

[1] G. B. Lesovik and L. S. Levitov, Phys. Rev. Lett. **72**, 538 (1994).
[2] S.-R. Eric Yang, Solid State Commun. **81**, 375 (1992).
[3] C. de C. Chamon, D. E. Freed, and X. G. Wen, Phys. Rev. B **51**, 2363 (1995).
[4] L. S. Levitov, H. Lee, and G. B. Lesovik, J. Math. Phys. **37**, 4845 (1996).
[5] L. D. Landau and E. M. Lifshitz, *Statistical Physics*, Pergamon Press, London, 1958.
[6] H. Nyquist, Phys. Rev. **32**, 110 (1928).
[7] See, for example: J. Perina, *Coherence of Light*, D. Reidel Publishing Co., Dordrecht, 1985.
[8] W. E. Lamb and R. C. Retherford, Phys. Rev. **72**, 339 (1947).
[9] H. B. G. Casimir, Proc. Kon. Ned. Akad. Wet. **51**, 793 (1948).
[10] L. Mandel, Phys. Rev. **152**, 438 (1966).
[11] R. Landauer, IBM J. Res. Dev. **32**, 306 (1988); D. S. Fisher and P. A. Lee, Phys. Rev. **23**, 6851 (1981).